\documentclass[preprint2]{aastex6}

\begin{document}

\title[Carbon chain anions \& molecular growth]{Carbon chain anions and the growth of complex organic molecules in Titan's ionosphere}

\author{
R. T. Desai\altaffilmark{1,2}, 
A. J. Coates\altaffilmark{1,2}, 
A. Wellbrock\altaffilmark{1,2},
V. Vuitton\altaffilmark{3}, 
F. J. Crary\altaffilmark{4}, 
D. Gonz\'alez-Caniulef\altaffilmark{1}, 
O. Shebanits\altaffilmark{5,6},
G. H. Jones\altaffilmark{1,2} 
G. R. Lewis\altaffilmark{1}
J. H. Waite\altaffilmark{7}, 
S. A. Taylor\altaffilmark{1,2},
D. O. Kataria\altaffilmark{1} 
J. -E. Wahlund\altaffilmark{5,6}
N. J. T. Edberg\altaffilmark{5} 
and E. C. Sittler\altaffilmark{8}}
\altaffiltext{1}{Mullard Space Science Laboratory, 
University College London, Holmbury St. Mary, Surrey, RH5 6NT, UK.}\email{r.t.desai@ucl.ac.uk}
\altaffiltext{2}{Centre for Planetary Science at UCL/Birkbeck,London,   
 Gower Street, London WC1E 6BT, UK.}
\altaffiltext{3}{Universit\'e Grenoble Alpes, CNRS, IPAG, F-38000 Grenoble, France.}
\altaffiltext{4}{Laboratory for Atmospheric and Space Physics, University of Colorado, Innovation Dr, Boulder, CO 80303, USA.}
\altaffiltext{5}{Department of Physics and Astronomy, Uppsala University, Box 516, SE 751 20 Uppsala, Sweden.}
\altaffiltext{6}{Swedish Institute of Space Physics, Box 537, SE 751 21 Uppsala, Sweden.}
\altaffiltext{7}{Space Science and Engineering Division, Southwest Research Institute (SWRI), 6220 Culebra Road, San Antonio,
TX 78238, USA.}
\altaffiltext{8}{NASA Goddard Space Flight Center, 8800 Greenbelt Road,
Greenbelt, MD 20771, USA.}

\accepted{June 01, 2017}

\begin{abstract}
 
Cassini discovered a plethora of neutral and ionised molecules in Titan's ionosphere including, surprisingly, anions and negatively charged molecules extending up to 13,800~u/q. In this letter we forward model the Cassini electron spectrometer response function to this unexpected ionospheric component to achieve an increased mass resolving capability for negatively charged species observed at Titan altitudes of 950-1300~km. We report on detections consistently centered between 25.8-26.0~u/q and between 49.0-50.1~u/q which are identified as belonging to the carbon chain anions, CN$^-$/C$_3$N$^-$ and/or C$_2$H$^-$/C$_4$H$^-$, in agreement with chemical model predictions.  At higher ionospheric altitudes, detections at 73-74~u/q could be attributed to the further carbon chain anions C$_5$N$^-$/C$_6$H$^-$ but at lower altitudes and during further encounters, extend over a higher mass/charge range. This, as well as further intermediary anions detected at $>$100~u, provide the first evidence for efficient anion chemistry in space involving structures other than linear chains. Furthermore, at altitudes below $\sim$1100~km, the low mass anions ($<$150~u/q) were found to deplete at a rate proportional to the growth of the larger molecules, a correlation that indicates the anions are tightly coupled to the growth process. This study adds Titan to an increasing list of astrophysical environments where chain anions have been observed and shows that anion chemistry plays a role in the formation of complex organics within a planetary atmosphere as well as in the interstellar medium.

\end{abstract}

\keywords{  Planets and satellites: individual (Titan) - Planets and satellites: atmospheres - Planets and satellites: composition - Astrochemistry - Astrobiology - ISM: molecules}

\section{Introduction} \label{sec:intro}

	Titan is the second largest moon in the Solar System (radius $R_T$ = 2576 km) and possesses a dense extended atmosphere principally composed  of $\sim$96\% molecular nitrogen, $<$4\% methane, and $<$1\% hydrogen \citep{Vervack04,Niemann05,Waite05}. Aerosol-type particles envelop the moon in a thick organic photochemical haze \citep{Danielson73}, a phenomenon also present at Pluto \citep{Gladstone16},  Triton \citep{Broadfoot89}, the Archean Earth \citep{Miller59} and likely also methane rich extra-solar planets. The production mechanisms and composition of these naturally occurring organic compounds are however far from understood. 
	
	The Cassini spacecraft has sampled the ionised regions of Titan's upper atmosphere down to altitudes of $<$900~km in-situ and observed positively charged ions (cations) extending up to nearly 1000~u/q \citep{Crary09,Coates10} and, surprisingly, negatively charged ions (anions) and aerosol precursors extending up to $13,800$~u/q \citep{Waite07,Coates07,Coates09}. The cations were detected at nearly all masses up to 100~u with over 50 species identified in this range \citep{Cravens06,Vuitton07}. At $>$100 u, evidence for carbon-based aromatic compounds has been reported, although unique identifications were not possible \citep{Crary09, Westlake14, Wahlund09}. The anions and larger negatively charged molecules were obtained at a lower resolution and classified into broad mass groupings of 12-30, 30-55, 55-90, 90-125, 125-195, 195-625, and 625$+$ u/q with the higher masses observed at lower altitudes and higher latitudes \citep{Coates07,Wellbrock13}. In the deep ionosphere below $\sim$1000~km the anion/aerosol precursor charge density was observed to exceed that of the electrons resulting in an ion-ion (dusty) plasma \citep{Shebanits13,Shebanits16}.
	
	\begin{figure*}
\includegraphics[width=0.85\textwidth]{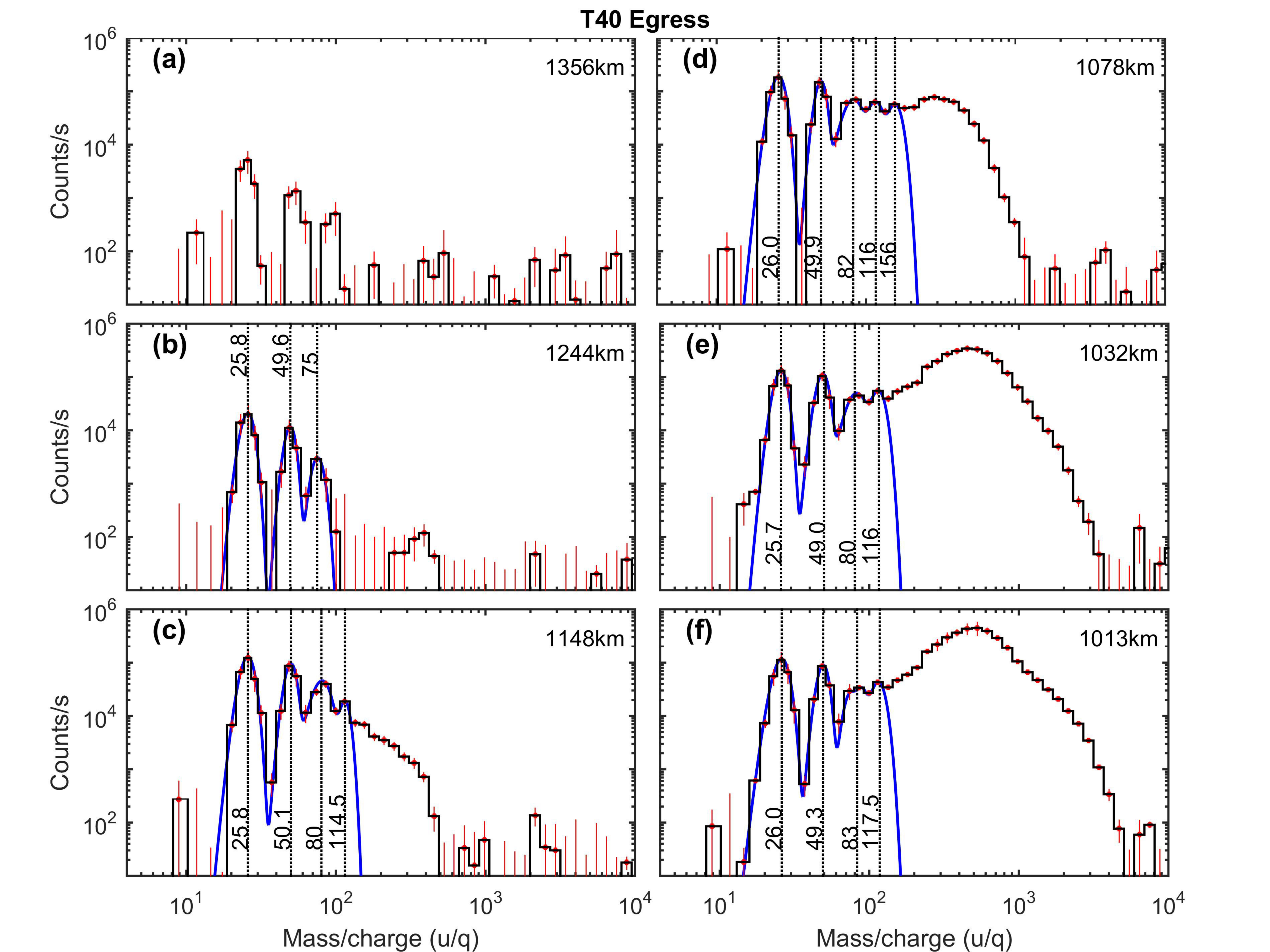}
\centering
\caption{Histogram of the CAPS-ELS anion ion mass/charge spectrum at various altitudes within Titan's ionosphere during the T40 encounter. The fitting routine (blue) and errorbars (red) are calculated as described in Equations \ref{eq1}-\ref{eq7}. The nominal centre of each group is marked (dotted black line) and the fitting parameters are given in Table  \ref{table1}. 
\label{fig2}}
\end{figure*}
	
	A number of studies have modelled the cation and neutral chemistry occurring in Titan's ionosphere at $<$100~u/q, but only two studies have attempted to model the anion chemistry \citep{Vuitton09,Dobrijevic16}. These focussed on low mass species of $<$75~$\textrm{u}$ and in particular 
inferred the presence of the C$_{n-1}$N$^-$ and C$_{n}$H$^-$ anions, where n=2-6. These carbon chain anions have all also been detected in  dark molecular clouds, prestellar
cores or protostellar envelopes \citep{Cordiner13,Millar17} where their high reactivity acts as a catalyst for the formation of larger organic molecules \citep{Millar00,Walsh09}. Chemical models of these environments also predict even larger anions containing up to 23 carbon atoms \citep{Bettens96}.

      At Titan, chemical schemes are only beginning to provide theories as to how the larger species can be produced. Stochastic charging models provide some explanation for how species of $\sim$100~u could be ionised and aggregate to form $>$10,000~u molecules \citep{Michael11, Lavvas13, Lindgren16}  but only a few studies have looked at precise chemical routes for producing molecules $>$100 u.
	 \citet{Westlake14} demonstrate how cations of $>$100~u are likely formed from smaller hydrocarbon compounds through ion-molecule growth processes and \citet{Ali15} provided a mechanistic analysis of possible routes from small to large cations of $<$250~u/q. The latter was based upon Olah's three-membered H\"uckel aromatic rings \citep{Olah72, Olah16}, and suggest the presence of several carbocations and their corresponding carbanions.  
	
	This letter provides analysis of the anion and aerosol precuror dataset using a forward model of the Cassini Plasma Spectrometer (CAPS) Electron Spectrometer (ELS) instrument response function to these species. Statistical evidence for the low mass carbon chain anions, CN$^-$/C$_2H^-$ and C$_3$N$^-$/C$_4$H$^-$ is presented as well as constraints on intermediary anions in the range of 50-200~u/q. The role of these species is then examined with respect to the growth of larger molecules with decreasing altitude.

\section{Methodology} \label{sec:ps1_data}

The results presented in this letter are derived from CAPS-ELS observations obtained during the T16, T18, T32, T40 and T48 encounters. These include measurements across Titan's sunlit (T16 ingress, T18, T32 ingress, T40, T48) and anti-sunlit (T16 egress \& T32 egress) hemispheres, a variety of latitudes and also when Titan was immersed directly within the solar wind (T32). 
	Further information on the geometry and ambient conditions of the Titan encounters can be found in \citet{Coates09}. 
	
	The CAPS-ELS is a top-hat electrostatic-analyser sensitive to negatively charged particles in the 0.6 - 28,000 eV range \citep{Young04}. Anions can be identified within the ELS three-dimensional velocity distribution due to being highly super-sonic in the spacecraft frame and  preferentially registering in anodes aligned with the spacecraft velocity vector. Thus as Cassini travels through Titan's ionosphere, the ELS observes an anion mass/charge spectrum,
	
\begin{equation}
\label{eq1}
\frac{m}{q}=\frac{2}{qv_{sc}^2}\left(E_{ELS}-\phi_{sc}\right)
\end{equation}	
where $E_{ELS}$ is the nominal acceptance energy, $v_{sc}$ is the spacecraft velocity relative to Titan and $\phi_{sc}$ is the ELS spacecraft potential shift applied in accordance with Liouville's theorem \citep{Lewis08}. 

	To isolate the anion detections, the count rates are taken from each scan across the ram direction and the isotropically observed electrons on non-ram pointing anodes averaged and subtracted. 
 The count rate $R_C$, can then be related to the number density, $n_{ni}$, using the ion current approximation \citep{Waite07,Coates07},
\begin{equation}
\label{eq2}
n_{ni}=\frac{R_c}{  v_{sc} \hspace{0.5mm}  A_F \hspace{0.5mm} \varepsilon_{} },
\end{equation}
where $A_F = 0.33~\textrm{cm}^2$ is the effective area of acceptance, and $\varepsilon_{}$ is the Microchannel Plate (MCP) anion detection efficiency function which is energy dependent. A value of $\varepsilon_{}=0.05$ was used in previous studies based upon the extensive study by \citet{Fraser02} which remains the best estimate for the larger species. Studies have however shown that at lower energies this could be significantly larger for negatively charged molecules \citep[e.g.][]{Peko00} where electron multiplication is increasingly dependent on potential as well as kinetic emission processes \citep{Hagstrum88}. Further analysis of the density uncertainties will be addressed by Wellbrock et al. (manuscript in preparation) and in this study we do not implement a new value.

	 The ELS energy bins are quasi-logarithmically spaced to match the energy resolution and overlap at FWHM which results in a nominal electron or anion distribution registering counts across multiple energy bins. To take this into account we forward model the ELS response to negative ions. The ELS $\Delta E/E$ resolution can be represented using a normalised Gaussian of the form
	 
\begin{equation}
\label{eq3}
f(E) =  R_{nc} \exp \left( -\frac{1}{2}\left( \frac{E-E_0}{E_W} \right )^2 \right ) 
\end{equation}
where $E_0$ is the center of the distribution, $E_W$ is the width and $R_{nc}$ is the normalised count rate. Here, $E_W$ corresponds to $\Delta E/E=16.7\%$ in the case of a single anion distribution but can be larger in the case of multiple overlapping distributions. The thermal energy spread of the anions is approximated by a drifting Maxwellian expressed in count rates,

\begin{equation}
\label{eq4}
g(E)=\frac{2nE^2G}{\sqrt{m}}\left(\frac{1}{2 \pi kT}  \right)^{\frac{3}{2}} \exp \left(-\frac{(E-E_0)}{kT}\right),
\end{equation}
as adapted from \citet{Rymer01}, where $T$ is the ion temperature in Titan's ionosphere and $G$ the geometric factor, 
		\begin{equation}
\label{eq5}
G=A_F \frac{\Delta E}{E}  \varepsilon,
\end{equation}  
which is derived under the assumption that the negative ion current fills the ELS aperture. The ion temperature in Titan's ionosphere has been determined to be significantly less than the $\Delta E/E$ instrument sensitivity and tests with or without this thermal contribution produced similar results. In this study it is therefore held constant at $kT = 0.02$~eV ($\sim$150K) \citep{Crary09} and included for completeness. 
  The ELS response function and the anion distribution can then be convolved, 
		\begin{equation}
\label{eq6}
h(E)= f(E) \ast g(E),
\end{equation}  
and the resulting function modelled to fit the observed data using a ${\chi}^2$ minimisation routine.

	 Of further mention is that the spacecraft surfaces charge to negative values in Titan's relatively dense ionosphere and the various surfaces will also charge to different potentials based upon variations in material conductivities and incident electron and ion currents \citep{Crary09}. This results in the exact potential correction, $\phi_{sc}$, also being unknown and the centers of the fitted distributions are therefore established relative to one another. Errors in the observed count rates are taken as

\begin{deluxetable*}{ccccccccccccc}
\tablenum{1}
\tablecaption{Fitting results for peaks 1-5 in the CAPS-ELS  anion mass/charge spectrum at $<$200 u/q as marked in Figure \ref{fig2} and Figure \ref{fig3}. The $\chi^2_{red}$ values are for 3 DOF fits within the 20-60 u/q range and the super- and sub-scripts correspond to a 2$\sigma$ deviation. Here, notice that for $p=0.05$ we obtain a critical $\chi^2_{red} =2.60$. The ELS FWHM for a single distribution function is $\sim$16.7$\%$ and the spacecraft potentials correspond to the nearest RPWS-LP measurement of this parameter to closest approach.
\label{table1}}
\tablewidth{0.1pt}
\tablehead{
\colhead{} & \colhead{} & \colhead{} &
\multicolumn{2}{c}{\textbf{Peak 1}}                                                                                                & \multicolumn{2}{c}{\textbf{Peak 2}}                                                                                                
& \colhead{\textbf{Peak 3}} & \colhead{\textbf{Peak 4}} & \colhead{\textbf{Peak 5}} & \multicolumn{2}{c}{\textbf{Potential ($\mathbf{\phi_{sc}}$})}                                                                                                 \\
\colhead{\textbf{Flyby}} & \colhead{\textbf{Alt.}}  & \colhead{\boldmath$\mathbf{\chi^2_{red}}$} &
 \colhead{\textbf{Center}} & \colhead{\textbf{FWHM}} &  \colhead{\textbf{Center}} & \colhead{\textbf{FWHM}} &
  \colhead{\textbf{Range}} &  \colhead{\textbf{Range}} &  \colhead{\textbf{Range}} &
   \colhead{\textbf{RPWS-LP}} &  \colhead{\textbf{ELS}} 
   \\ 
   \colhead{$\mathbf{(\#)}$} & \colhead{$\mathbf{(km)}$}  & \colhead{$\mathbf{(-)}$} &
 \colhead{$\mathbf{(u/q)}$} & \colhead{\boldmath${(\%)}$} &  \colhead{$\mathbf{(u/q)}$} & \colhead{\boldmath${(\%)}$} &
  \colhead{$\mathbf{(u/q)}$} &  \colhead{$\mathbf{(u/q)}$} &  \colhead{$\mathbf{(u/q)}$} &
   \colhead{$\mathbf{(V)}$} &  \colhead{$\mathbf{(V)}$} 
  }
\startdata
\textbf{T40} & 1244 & $\rightarrow$0 & $25.8_{-0.6}^{+0.8}$ & $20.1_{-4.6}^{+2.6}$ & $49.6_{-1.5}^{+1.8}$ & $20.9+$ & 72-78 & - & - & -0.63 & -1.05
\\
\textbf{-} & 1148 & 1.50 & $25.8_{-0.5}^{+0.9}$ & $21.7_{-3.3}^{+2.1}$ & $〖50.1〗_{-0.8}^{+1.1}$ & $〖17.9〗_{-2.7}^{+1.5}$ & 71-89 & 109-120 & - & -0.59 & -1.20
\\
\textbf{-} & 1078 & 0.67 & $26.0_{-0.6}^{+0.4}$ & $〖21.6〗_{-2.8}^{+0.6}$ &$〖49.9〗_{-1.1}^{+2.0}$ &  $〖17.0〗_{-3.5}^{+2.2}$ & 73-91 & 108-122 & 146-166 & -0.58 & -1.20
\\
\textbf{-} & 1032 & 0.72 & $25.7_{-0.5}^{+0.5}$ & $〖22.7〗_{-2.0}^{+0.6}$ & $〖49.0〗_{-0.5}^{+0.8}$ & $〖19.6〗_{-2.0}^{+1.4}$ & 73-91 & 109-123 & - & -0.57 & -1.20
\\
\textbf{-} & 1013 & 2.46 & $26.0_{-0.6}^{+0.8}$ & $〖22.7〗_{-2.9}^{+1.9}$ & $〖49.3〗_{-0.6}^{+0.6}$ & $〖18.1〗_{-1.6}^{+1.0}$ & 72-94 & 107-128 & - & -0.59 & -1.20 
\\
\textbf{T16} & 1031 & 0.88 & $25.9_{-0.6}^{+0.4}$ & $〖22.8〗_{-6.0}^{+2.5}$ & $〖48.9〗_{-0.7}^{+0.7}$ & $〖19.6〗_{-2.2}^{+1.6}$ & 72-77 & 108-118 & 141-160 & -0.66 & -1.40
\\
\textbf{T18} & 969 & 0.73 & $26.0_{-1.0}^{+0.9}$ & $〖22.8〗_{-6.2}^{+3.7}$ & $〖49.0〗_{-0.7}^{+1.5}$ & $〖17.0〗_{-2.1}^{+2.0}$ & 74-82 & 113-123 & 148-166 & -1.65 & -2.50  
\\
\textbf{T32} & 1036 & 0.92 & $25.8_{-1.1}^{+0.6}$ & $〖18.5〗_{-2.4}^{+3.7}$ & $〖49.5〗_{-1.2}^{+0.8}$ & $〖18.5〗_{-3.5}^{+1.8}$ & 74-94 & 114-125 & - & -0.80 & -1.55
\\
\textbf{T48} & 1082 & 2.18 & $26.0_{-0.2}^{+0.2}$ & $〖18.3〗_{-0.4}^{+0.7}$ & $〖49.7〗_{-0.6}^{+0.7}$ & $〖17.7〗_{-1.8}^{+1.1}$ & 74-88 & 110-121 & 147-164 & -0.80 & -1.40
\\
\enddata
\end{deluxetable*}	
	
\begin{equation}
\label{eq7}
\sigma = \sigma_p + \sigma_{std}
\end{equation}  
where $\sigma_{p}$ corresponds to the Poisson counting statistics and $\sigma_{std}$ corresponds to  the standard deviation of counts on non-ram oriented anodes. This is used as a measure of electron anisotropies and inter-anode scaling uncertainties, introduced when isolating the anion detections.

\section{Species identification} \label{sec:Carbon chain identification}

\begin{figure*}
\centering
\includegraphics[width=0.85\textwidth]{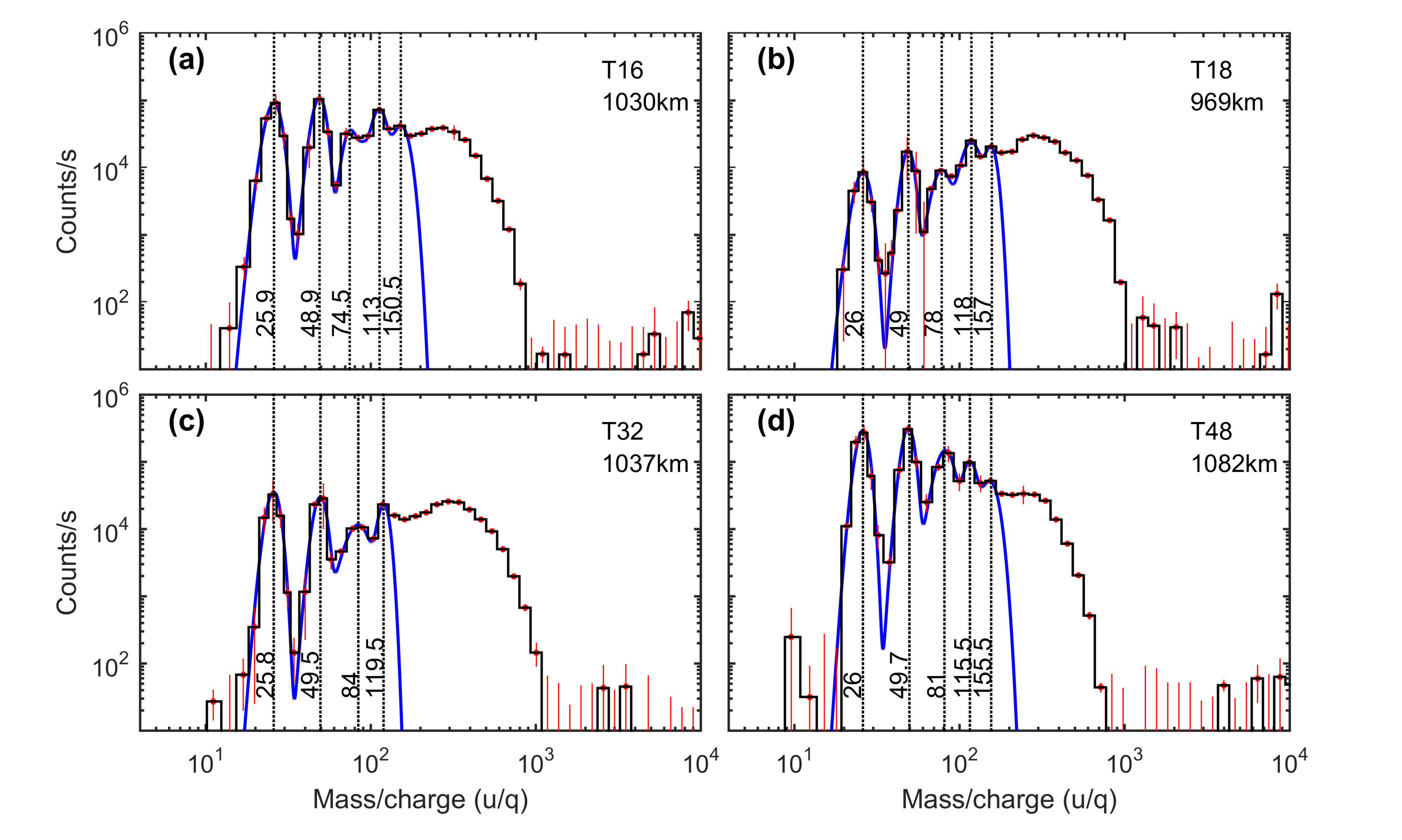}
\caption{Histogram of the CAPS-ELS anion mass/charge spectrum during the T16, T18, T32 and T48 encounters. The fitting routine (blue) and errorbars (red) are calculated as described in Equations \ref{eq1}-\ref{eq7}.  The nominal centre of each group is marked (dotted black line) and the fitting parameters provided in Table \ref{table1}.
\label{fig3}}
\end{figure*}

	Figure \ref{fig2} shows the anion mass/charge spectrum measured at various altitudes during T40. At higher altitudes ($>1300$ km) the larger $>$100 u/q species are absent and at the highest altitudes it is difficult to identify anions due to decreased densities. As Cassini descends, clear detections appear at $<$200 u/q and the larger $>$200 u/q distribution starts to grow below $\sim$1250 km. The five resolved clustered detections in the spectra are hereafter referred to as peaks 1-5 and fall within the range of mass groups 1-5 as described by \citet{Wellbrock13}. Figure \ref{fig3} also shows anion spectra obtained during encounters T16, T18, T32 and T48. 
		
		The fitting procedure applied to peaks 1 and 2 finds the center of the primary two peaks to be  separated by $23-24.3$~u/q in all encounters, see Table \ref{table1}. Chemical models for Titan’s atmosphere predict efficient production of C$_n$H$^-$ and C$_{n-1}$N$^-$ to result from dissociative electron attachment to, or de-protonation of, parent neutral species C$_n$H$_2$ and HC$_{n-1}$N \citep{Vuitton09,Dobrijevic16}. For example, CN$^-$ and C$_3$N$^-$ are produced by
			\begin{equation}
\label{eq8}
 {HCN + e^- \rightarrow H + CN^-},
\end{equation}  
			\begin{equation}
\label{eq9}
HC_3N + CN^-  \rightarrow  HCN + C_3N^-,
\end{equation}  
which proceeds rapidly due to abundant HCN and HC$_3$N. A similar reaction sequence exists for hydrocarbons where C$_2$H$^-$ and C$_4$H$^-$ are produced by
 			\begin{equation}
 \label{eq10}
 {C_2H_2 + e^- \rightarrow H + C_2H^-,}
\end{equation}  
			\begin{equation}
\label{eq11}
{C_4H_2 + C_2H^- \rightarrow C_2H_2 + C_4H^-.}
\end{equation}  
due to abundance C$_2$H$_2$ and C$_4$H$_2$. This $23-24.3$~u/q separation in the ELS mass/charge spectrum is indicative of these processes and the CN$^-$/C$_2$H$^-$ and C$_3$N$^-$/C$_4$H$^-$ carbon chain anions as the dominant constituents within the primary and secondary peaks respectively. It is not possible to further resolve the 1~u difference between these nitrile and hydrocarbon compounds but CN$^-$ is estimated to be two orders of magnitude more abundant than C$_2$H$^-$, and C$_3$N$^-$ and C$_4$H$^-$ are predicted in comparable abundances \citep{Vuitton09}. The main anion loss process considered is associative detachment with neutral radicals.

		 The width of the primary peak is however often larger than the ELS $\Delta E/E\approx16.7\%$. This can be explained by a multi-species composition, with further possible anion species such as C$_2^-$, CH$_{1,2,3}^-$, NH$_{1,2,3}^-$ and CO$^-$ possibly contributing, the latter due to the introduction of water-group ions (O$^+$, OH$^+$, H$_2$O$^+$, H$_3$O$^+$) from Enceladus \citep{Hartle06,Cravens08}. It is however possible that the spacecraft potential also acts to spread a given distribution's energy relative to the spacecraft, an effect which would be more pronounced for lower mass ions due to their lower inertia.  The width of the secondary C$_3$N$^-$/C$_4$H$^-$ peak falls across a range which encompasses the FWHM of the ELS, indicating this is likely composed of a single distribution function with only a minor contribution from further species possible. 
		 
		 This analysis indicates the spacecraft potential experienced by the ELS is $0.4-0.9$~V more negative than that measured by the Radio and Plasma Wave Science (RPWS) Langmuir Probe (LP) although within the $-3.5$~V absolute range observed by the instrument in Titan's ionosphere \citep{Crary09}. An $\sim$-0.3~V discrepancy was found between conjugate CAPS Ion Beam Spectrometer (IBS) and Cassini's Ion and Neutral Mass Spectrometer (INMS) observations \citep{Crary09} and a more negative spacecraft potential correction is expected for anion detections \citep{Jones11}.  This is due to focussing effects where the spacecraft-generated potential field acts to deflect incident ions such that the anions arrive from a direction closer to the spacecraft surface. 

\begin{figure*}
\centering
\includegraphics[width=1.0\textwidth]{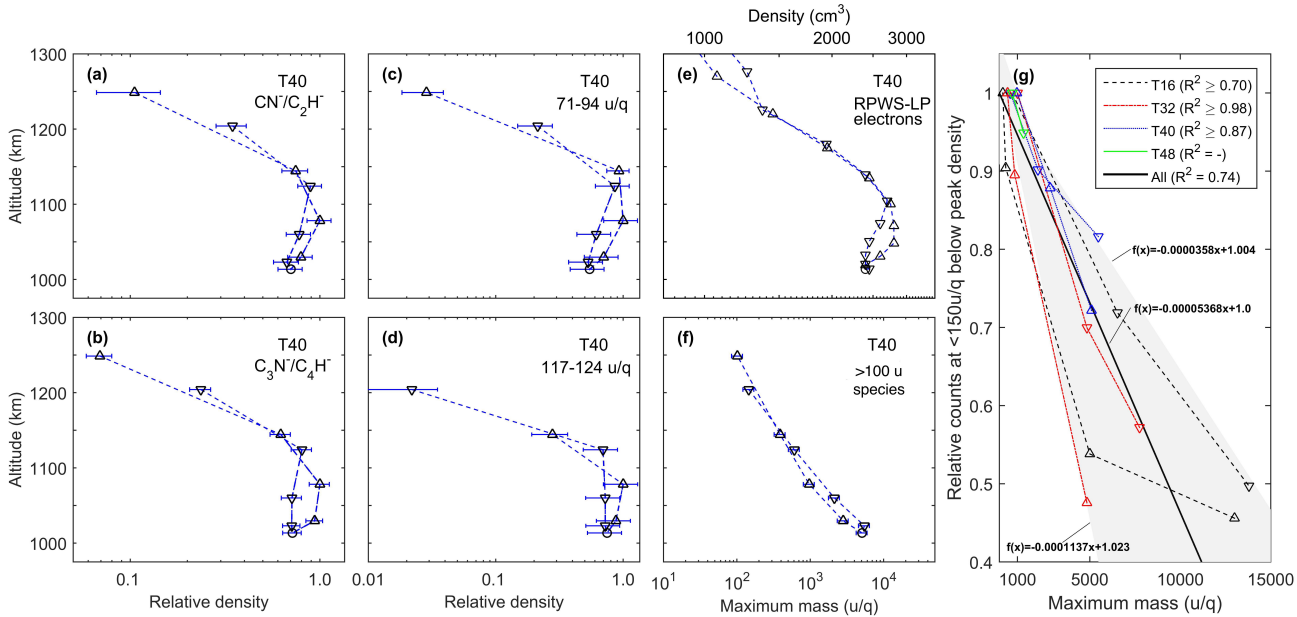}
\caption{Altitude profiles with ingress ($\triangle$) and egress ($\triangledown$) marked. (a-d) Shows the T40 relative anion densities, (e) the electron density, (f) the maximum mass detected and (g) the correlation between the depletion of the low mass ($<150$~u/q) anions below the altitude of peak density and corresponding increase in the maximum mass. A linear trend line (black) is fitted as well as individually to the ingress and egress of each encounter, the  maximum and minimum of which encompass the region shaded grey. Relative errors are assumed small compared with the overall spread. 
\label{fig4}} 
\end{figure*}

At 1244 km (Figure \ref{fig2}b) the third peak is also visible where at higher altitudes it appears compatible with C$_5$N$^-$/C$_6$H$^-$, although this cannot be statistically verified as at $>$50~u/q there are fewer measurements than free parameters. At lower altitudes this peak widens, extending as high as $\sim$94~u/q in some instances. 
	This range does not include any previously observed anions and indicates the presence of anionic structures other than linear chains. 
	INMS and CAPS-IBS measurements at these altitudes also show a grouping of neutrals and cations over a similar range \citep{Waite07,Crary09}, the most abundant of which was inferred to be benzene (C$_6$H$_6$) \citep{Vuitton08}. This is particularly relevant as benzene and benzene products are thought to be the seeds for larger aromatic compounds \citep{Vuitton16}. The lack of reaction rates and known chemical pathways at these high masses severely restricts the analysis of the anion chemistry but several candidate species can be suggested. For example, the cyclic radical C$_6$H$_5^+$ is observed in appreciable quantities \citep{Vuitton16} and can be a source for C$_6$H$_5^-$ anion radical production \citep{Fenzlaff84}. The stable C$_6$H$_7^-$ cyclic anion can also be produced through the interaction of benzene with H$^-$ \citep{Coletti12}. H$^-$ cannot be measured however due to constraints imposed by the spacecraft velocity (see Equation \ref{eq1}) but H$^-$ fluxes can result from the interaction of methane with ionospheric electrons \citep{Dobrijevic16}. 
	
	The four further encounters in Figure \ref{fig3} also show this third peak to be the most variable between encounters, possibly indicating an enhanced sensitivity to the ambient conditions. A number of azine anions resulting from benzene, pyridine, pyridazine, pyrazine, and s-triazine have been explored by \citet{Wang15} with application to Titan, and determined to be highly reactive with nitrogen and oxygen. These include C$_6$H$_3^-$, C$_5$H$_2$N$^-$, C$_5$H$_3$N$^-$, C$_5$H$_4$N$^-$, C$_4$H$_3$N$_2^-$ and C$_3$H$_2$N$_3^-$ and are of high astrobiological interest. The higher end of this third mass range is however less explored and further anions such as the benzyl anion C$_7$H$_7^-$ and anilinide anion C$_6$NH$_6^-$ \citep{Wang16}, could be derived from the $>20$ mostly Polyaromatic Hydrocarbons and nitrated heterocyclic species suggested in this range. 	

At altitudes $<$1200~km a fourth distinct peak, with consistent detections at $\sim$117$\pm$3~u/q, is identified during every encounter. This appears at count rates comparable to the primary peaks and sometimes double that of the neighboring peaks. This is most evident during T16 where the highest mass species at 13,800~u/q were observed \citep{Coates09}. A smaller fifth peak, with detections consistently at $\sim$154$\pm$8~u/q, is also sometimes present. The $>$100~u/q regime is however even more unconstrained. It should be noted though that \citet{Trainer13} detected ring structures near 117~u in laboratory simulations of aerosol-tholin production which could represent growth processes involving aromatic rings at Titan.
	   
	Multiply charged anionic states are not considered here due to inter-electron repulsive forces making this phenomena increasingly unlikely for smaller molecules. This is evident as the smallest known multiply charged anions, C$_n^{2-}$  (n=7-28), have lifetimes of tens of micro-seconds in the gas phase whereas the larger C$_{60,70}^{2-}$ molecules can persist for milliseconds  \citep[][]{Wang09}. It therefore appears that multiple charges are much more likely on molecules larger than a few hundred amu in Titan's ionosphere, as indeed was reported by \citet{Shebanits16}. 
	
\section{Molecular growth}

	 	Figure \ref{fig4} (a-d) shows the altitude profiles of peak 1 (25.8-26.0~u/q, associated with CN$^-$/C$_2$H$^-$), peak 2 (49.0-50.1~u/q, associated with C$_3$N$^-$/C$_4$H$^-$), peak 3 (71-94~u/q), and peak 4 (107-123~u/q) during the T40 encounter. Figure \ref{fig4} (e-f) shows the electron densities and maximum detected mass of the larger molecules as obtained in \citet{Coates09}. The altitude profiles in Figure \ref{fig4} show the carbon chain anions to peak in density above the region where the highest mass aerosol monomer are observed and to extend several scale heights above this to where the larger species are not present in measurable quantities. While the precise relative densities are not known, the profiles show the negative charge to be increasingly carried by the larger species at lower altitudes. This trend is also observed in further encounters, see \citet{Wellbrock13}, Figure 3.
	   	
  Below $\sim$1100~km the depletion of the $<$150~u/q anions can be seen to be related to the size increase of the larger molecules. Figure \ref{fig4} (g) shows this correlation for all encounters except T18 which was omitted as the altitude of peak density of the low mass ions was not definitively surpassed (indeed during T48 this was only surpassed for a brief instance). The two parameters can be seen to be linearly proportional although there is some spread to the data at the higher masses. 	   	     This overall proportional decrease in the low mass species with the increase of the larger species points to dependencies between these.

   The data also point to a possible diurnal variation with the day-side measurements grouped together at and above the fitted trend line (black) and the night-side measurements appearing below this and with T16 and T32 crossing Titan's solar terminator. \citet{Coates09} previously determined significant spatial variations of the high mass monomers and these data point to this being echoed in the smaller species. Further statistical analyses are however required to fully disentangle such influences.
   
\section{Summary \& Conclusions} \label{sec:cite}

	This study used a forward model of the CAPS-ELS response function to achieve an increased mass resolving capability for anions  in Titan ionosphere, the results of which are as follows: The first peak ($25.8-26.0$~u/q) and second peak ($49.0-50.1$~u/q) in the anion spectrum were statistically shown to be compatible with the CN$^-$/C$_2$H$^-$ and C$_3$N$^-$/C$_4$H$^-$ respectively, although it was not possible to differentiate between these nitrile or hydrocarbon compounds. At altitudes above $\sim$1200~km the third peak is consistent with the further chain anion C$_5$N$^-$/C$_6$H$^-$ but at lower altitudes becomes dominated by higher mass species not consistent with carbon chain anions. Notably, this is evidence for the presence of more complex structures which may well come to represent the first astrophysical detection of anions not composed of linear chains, see recent review by \citet{Millar17}. A number of species were suggested to account for this intermediary peak based upon the relatively better understand cation and neutral chemistry. Further persistent detections were also constrained at 117$\pm3$~u/q and 154$\pm$6~u/q which appear to also add to the growing body of evidence for growth processes involving ring structures although the current lack of known reaction pathways impedes their definitive interpretation at present. 
	
	The evolution of the low mass ($<$150~u/q) anions was then examined with respect to the impending growth of the larger organic molecules. Deep within the ionosphere the lower mass anions were observed to become depleted as the larger aerosol precursors coincidentally underwent rapid growth. This trend contributes to the idea that smaller species form the seeds for the larger species via a series of reactions and processes which the chain anions and further intermediary anions appear to be tightly coupled to. These results demonstrate the importance of tracing a route from small to large species in order to fundamentally understand how complex organic molecules can be produced within a planetary atmosphere.    	

\acknowledgments
RTD acknowledges STFC Studentship 1429777. AJC, AW and GHJ acknowledge support from the STFC consolidated grants to UCL-MSSL ST/K000977/1 and ST/N000722/1. DGC acknowledges Becas-Chile CONICYT Fellowship (No. 72150555). OS acknowledges funding from SNSB, Dnr 130/11:2

\end{document}